\documentclass[aps,prd,floats,nofootinbib,preprintnumbers,11pt]{revtex4}
\usepackage{epsfig}
\usepackage{amsmath}
\usepackage{amssymb}
\usepackage{amsthm}
\usepackage{longtable}
\usepackage{array}
\usepackage[T1]{fontenc}
\usepackage[latin9]{inputenc}
\usepackage{color}
\usepackage{graphicx}
\usepackage{esint}

\begin{document}

\preprint{LA-UR-08-06358}
\preprint{NUHEP-TH/08-07}

\vspace{3cm}

\title{Minimally Allowed $\beta\beta0\nu$ Rates From Approximate Flavor
Symmetries}

\author{James Jenkins}

\email{jjenkins6@lanl.gov}

\affiliation{Theoretical Division, Los Alamos National Laboratory, Los Alamos, New Mexico~87545, USA}
\affiliation{Northwestern University, Department of Physics \& Astronomy, Evanston, IL~60208, USA}

\begin{abstract}
Neutrinoless double beta decay ($\beta\beta0\nu$) is among the only realistic probes of Majorana neutrinos.  In the standard scenario, dominated by light neutrino exchange, the process amplitude is proportional to $m_{ee}$, the $e-e$ element of the Majorana mass matrix.  Naively, current data allows for vanishing $m_{ee}$, but this should be protected by an appropriate flavor symmetry.  All such symmetries lead to mass matrices inconsistent with oscillation phenomenology.  I perform a spurion analysis to break all possible Abelian symmetries that guarantee vanishing $\beta\beta0\nu$ rates and search for minimally allowed values.  I survey 230 broken structures to yield $m_{ee}$ values and current phenomenological constraints under a variety of scenarios.  This analysis also extracts predictions for both neutrino oscillation parameters and kinematic quantities.  Assuming reasonable tuning levels, I find that $m_{ee}>4\times 10^{-6}~{\rm eV}$ at $99\%$ confidence.  Bounds below this value might indicate the Dirac neutrino nature or the existence of new light (eV-MeV scale) degrees of freedom that can potentially be probed elsewhere.
\end{abstract}
\maketitle

\section{Introduction\label{sec:introduction}}

Neutrino oscillation experiments have given conclusive evidence that
neutrinos have mass and mix. This constitutes the first terrestrial
evidence of physics beyond the Standard Model (SM) and leads to many
important questions.  For a recent review of neutrino physics see \cite{NuRev:TheoryWhitePaper,NuRev:AndreTASI,NuRev:AndreSoWhat}.
Broadly, it is puzzling why the neutral lepton sector is so different
from the other SM fermions in both mass scale and mixing pattern.
The resolution to this mystery has implications for both particle
and astrophysics and provides deep theoretical insight into the nature
of other high scale phenomena. The neutrality of the neutrino under
the only unbroken gauge symmetry is the likely key to this problem,
as it offers the possibility that the neutrino is its own antiparticle
via direct coupling within a Majorana mass term. Such a mass, as opposed
to the more common Dirac mass, is composed of only one field and violates
all non-zero quantum numbers by two units. The charge conjugation
properties of the neutrinos, their Dirac vs Majorana nature,
are currently unknown and their determination is arguably the most important
task facing the neutrino community.

The favored means of probing Majorana neutrinos is via the process
of neutrinoless double beta decay ($\beta\beta0\nu$) where, within
a nucleus, two neutrons decay into two protons with no neutrinos \cite{Elliott:DoubleBetaDecay}.
This process violates lepton number by two units and may proceed via
the virtual exchange of Majorana neutrinos. In this case, the decay
amplitude is directly proportional to the mass of the exchanged electron-type
neutrino, or more precisely the $e-e$ element of the Majorana neutrino
mass matrix ($m_{ee}$) taken in the flavor basis where the charged lepton masses are diagonal. Current experimental limits on the $^{76}$Ge isotope constrain the $\beta\beta0\nu$
half-life below $\sim10^{25}$ years, corresponding to $m_{ee}<0.35$
eV at 90\% confidence%
\footnote{The translation between measured half-life and $m_{ee}$ is not straightforward, as it depends critically on isotope dependent nuclear matrix element calculations, where uncertainties currently range within a factor of three
\citep{BB0nMatrixElUnc,Rodin:AssessmentUncertaintiesQRPABB0nNuclearMatrixElements,Menendez:DisassemblingNuclearMatrixElementsBB0n}.
This is likely to improve within the next several years. %
} \citep{IGEXBB0nResults,LatestResultsHeidelberMoscowBB0nExp,AbsoluteValuesNuMassStatusProspects,ImplicationsOfNuDataCirca2005}.
Next generation experiments are poised to extend this reach by roughly
an order of magnitude to $m_{ee}<0.05$ eV \citep{StrategiesForNextGenBB0nExperiments,AbsoluteValuesNuMassStatusProspects,Zuber:BB0nExperiments}. 

Of course, other exotic interactions can mediate $\beta\beta0\nu$,
but it was argued in the Blackbox theorem \citep{BlackBox,SUSYBlackbox}
that any such Lepton Number Violating (LNV) process will necessarily
yield a Majorana neutrino mass at some order in perturbation theory,
just as a Majorana neutrino mass term will lead to LNV processes.
This notion was extended in \citep{ExtendedBlackBox} to the realistic
three neutrino system. The authors showed, among other things, that there
exists a one-to-one relationship between LNV rates and elements of
the Majorana neutrino mass matrix such that, in particular, $m_{ee}=0\Longleftrightarrow\Gamma_{\beta\beta0\nu}=0$
and $m_{ee}\neq0\Longleftrightarrow\Gamma_{\beta\beta0\nu}\neq0$.
Additionally, using general symmetry arguments, they demonstrated
that there exists a non-trivial relationship between various mass matrix
elements, implying a finite set of textures with vanishing $m_{ee}$.
None of these are consistent with the observed oscillation data. This
leads naturally to the conclusion that if neutrinos are Majorana
particles, $\beta\beta0\nu$ must occur at a nonzero rate. 

Here, using similar logic, I explore exactly how small $m_{ee}$ can
be under a variety of circumstances. If light Majorana neutrino exchange
is the only mode of $\beta\beta0\nu$, this can be applied directly
to the interpretation of experimental results. The situation is not as straightforward in the face of other
contributions, as these will generally effect the mass and LNV rate
differently \citep{MyLNV}. Still, the one-to-one correspondence between
$m_{ee}$ and $\Gamma_{\beta\beta0\nu}$ adds confidence to the conclusion
that $m_{ee}$ is a good measure of the $\beta\beta0\nu$ rate for
small $m_{ee}$, even in the presence of arbitrary new physics. This
statement becomes exact in the limit of vanishingly small $m_{ee}$.

This analysis is conducted within the framework of Abelian flavor symmetries acting within a three light Majorana neutrino system.  It will become clear that the results can be extended beyond this paradigm to include non-Abelian groups.  Additionally, by the above argument, there is reason to believe that the qualitative existence of a lower $\Gamma_{\beta\beta0\nu}$ bound, as well as its connection to $m_{ee}$, should remain valid in the face of arbitrary new physics at scales down to approximately $100~{\rm MeV}$.  Beyond this point, new light degrees of freedom can contribute directly to the $\beta\beta0\nu$ system and restrict the result validity. Thus, limits that fall below extracted minimum
values are evidence for new light physics or the Dirac neutrino nature.  A measurement
near the derived lower bound would indicate a slightly broken symmetry mechanism
at work. 

It is natural to wonder how small $\Gamma_{\beta\beta0\nu}$ can be without the introduction of these flavor symmetry suppressions.  It is well known that, given current neutrino data, there is a well-defined range of allowed $m_{ee}$ values.  This depends on the neutrino mass hierarchy and oscillation parameters within the light Majorana neutrino exchange hypothesis. \cite{Bahcall:LearnFromBB0nExp,Petcov:TheoryProsBB0n,BB0nFutureNuOscPrecisionExp,MyNonOscHier,ImplicationsOfNuDataCirca2005}.  Neutrinos with normal mass spectra can yield vanishing $m_{ee}$ provided appropriate phase and parameter choices.  Next generation experiments will probe the quasi-degenerate and inverted hierarchy region of the allowed range \cite{Elliott:DoubleBetaDecay}.  Clearly, a positive measurement at this relatively large level would not indicate a flavor suppression of any kind.  The question becomes more involved as smaller values within the normal mass hierarchy are explored.  When is small $m_{ee}$ no longer an accident?  To answer this question, it is instructive to take the structure free limit and consider the neutrino mass matrix anarchy hypothesis \cite{Haba:2000be,Hall:1999sn}.  Here, the underlying neutrino model is sufficiently complicated such that the low energy mass matrix appears random and must be treated statistically.  In other words, the flavor basis is some random rotation from the mass eigenbasis.  An analysis of the allowed mixing angle distribution is straightforward and described in \cite{deGouvea:2003xe}.  However, there is an added level of ambiguity introduced whenever mass values are discussed, due to freedom in assigning an integration measure to the probability distributions.  These issues were studied in \cite{Haba:2000be,MyAnarchy}.  The distribution of $m_{ee}$ values within the anarchy scenario was surveyed in \cite{MyAnarchy} under a variety of conditions to conclude that $m_{ee}<5\times 10^{-3}~{\rm eV}$ implies the existence of a flavor symmetry mechanism, new light degrees of freedom, or the Dirac neutrino nature.  A $\beta\beta0\nu$ measurement above this limit could be attributed to either a flavor symmetry or random fluctuations of the neutrino mass matrix. Below the anarchy bound, the present analysis sets a limit on minimum $m_{ee}$ values and identifies what symmetries are responsible for the suppression.  Above the anarchy bound, it selects those broken symmetries that are allowed by the data and makes predictions for other observables that can further constrain the system.

This paper is organized as follows.  In Section \ref{sec:AbelianFlavorSymmetries}, I review Majorana neutrino masses and the status of current neutrino data used as constraints in the remainder of the analysis. I then introduce flavor symmetries in the context of the neutrino mass matrix and motivate the utility of Abelian groups as an ideal laboratory for a comprehensive search for minimal $\Gamma_{\beta\beta0\nu}$ rates.  In Subsection \ref{sub:UnbrokenAbelian}, I exhaustively enumerate the flavor symmetry structures that lead to exact $m_{ee}=0$ and explore their consequences for neutrino oscillation phenomenology.  These results are summarized in Table \ref{tab:AblSummary}, which illustrates that of the eleven possible symmetry classes, none are consistent with current data.  I break these flavor symmetries in Subsection \ref{sub:Broken-Abelian-Flavor} with the introduction of a single spurious $U(1)_f$ charged scalar field that acquires a real vacuum expectation value (vev).  Subsubsection \ref{subsub:Numerical-Results} forms the bulk of the analysis, where I numerically survey a comprehensive set of $230$ broken symmetry structures.  For each case (referred to loosely as models), I determine current constraint from data, extract the minimally allowed $m_{ee}$ values, and make predictions for future neutrino experiments.  Variations of these results, subject to improvements in future oscillation parameter measurements, are also studied.  I conclude in Section \ref{sec:Concluding-Remarks} with a summary of the results and a discussion of the limitations of the analysis in the face of new physics.

\section{Abelian Flavor Symmetries} \label{sec:AbelianFlavorSymmetries}

If lepton number is violated by physics at some high scale $\Lambda$, the effective low energy Majorana neutrino mass Lagrangian term may be written as
\begin{equation}
\mathcal{L}_\nu = \frac{1}{2}m_{\alpha\beta}\overline{\nu^c}_\alpha\nu_\beta.
\label{eq:MajMassLag}
\end{equation}
In the weak interaction basis where the charged leptons are diagonal, the Greek subscrips are flavor indices that run over the three generations $e$, $\mu$ and $\tau$.  The symmetric mass matrix $m_{\alpha\beta}$ may be diagonalized to yield positive real mass eigenvalues $m_1$, $m_2$ and $m_3$ by the neutrino mixing matrix in the PDG parametrization \cite{PDG06}
\begin{equation}
U = \left(\begin{array}{ccc}
c_{12}c_{13} & s_{12}c_{13} & s_{13}e^{-i\delta}\\
-s_{12}c_{23}-c_{12}s_{23}s_{13}e^{i\delta} & c_{12}c_{23}-s_{12}s_{23}s_{13}e^{i\delta} & s_{23}c_{13}\\
s_{12}s_{23}-c_{12}c_{23}s_{13}e^{i\delta} & -c_{12}s_{23}-s_{12}c_{23}s_{13}e^{i\delta} & c_{23}c_{13}
\end{array}\right)
\left(\begin{array}{ccc}
1 & 0 & 0\\
0 & e^{i\phi_2} & 0\\
0 & 0 & e^{i\phi_3}
\end{array}\right)
\end{equation} 
that describes the rotation from the flavor basis to the mass basis.  I use the shorthand $c_{ij} =\cos\theta_{ij}$ and $s_{ij}=\sin\theta_{ij}$ for notational convenience.  Due to symmetries
of the mixing matrix, the mixing angles may be constrained within
$\theta_{ij}\in(0,\pi/2)$ \cite{AndreAlex:Darkside,deGouvea:2008nm}with the Majorana and Dirac phases $\phi_j \in (0,\pi)$ and $\delta \in (0,2\pi)$ without loss of generality \cite{deGouvea:2008nm,Jenkins:RephasingInvariantsQuarkLeptonMixingMatrices}. By convention\footnote{See for example \cite{deGouvea:2008nm} for a summary of mass naming conventions and their relationship with other mixing parameters.}, the neutrino eigenstates are ordered in mass squared such that $m_1^2$, and $m_2^2$ have the smallest separation, the so-called solar mass squared splitting $\Delta m^2_S=m_2^1 - m_1^2 > 0$, while $m_3$ is the most distant state.  The identity of the lightest eigenstate depends on the mass ordering, with $\nu_1$ and $\nu_3$ existing as the lightest state for the normal and inverted hierarchies, respectively.  Neutrino oscillation data constrains the mass eigenvalue squared differences and mixing angles as can be seen from the first five entries of Table \ref{tab:Constraints} which lists both the current best fit values and $1\sigma$ uncertainties, as adapted from \cite{ThreeFlavorOscUpdate}.  See also \cite{StatusGlobFitsMaltoni07,ImplicationsOfNuDataCirca2005,Fogli:GlobalAnal3flavnuMassMix}. Additionally, kinematic probes of the endpoint of the tritium beta decay spectrum, cosmological observations and even $\beta\beta0\nu$ constrain the absolute neutrino mass values \cite{AbsoluteValuesNuMassStatusProspects} as shown in entries six, seven and eight.  To date, these have yet to observe positive signals but have been successful in bounding neutrino masses below the eV level \cite{ImplicationsOfNuDataCirca2005}.  Next generation experiments will extend the reach of tritium decay and cosmological measurements to $m_{\nu_e}<0.2 ~{\rm eV}$ \cite{TritBDecKatrinLOE} and $\Sigma <0.1 ~{\rm eV}$ \cite{CosmoSumClusterExt,CosmoSumLensing,CosmoSumRedshift}, at $90\%$ confidence, respectively.  The single Dirac phase $\delta$ and two Majorana phases $\phi_2$, and $\phi_3$ are currently unconstrained by experiment.  The first column of Table \ref{tab:Constraints} lists the parameter name conventions incorporated in this analysis.  The subscripts $S$, $A$ and $R$ attached to the oscillation parameters refer respectively to ``solar'', ``atmospheric'' and ``reactor,'' after the primary/historical neutrino sources used in their measurement.  These parameter constraints must be satisfied by all viable neutrino mass models. 

\begin{table}
\begin{tabular}{|c|c|c|c|}
\hline 
Name & Parameter Combination & Value & $1\sigma$ Uncertainty\tabularnewline
\hline
\hline 
$\Delta m^2_S$ & $m_2^2 - m_1^2$ & $7.65\times 10^{-5}~{\rm eV^2}$ & $0.22\times 10^{-5}~{\rm eV^2}$\tabularnewline
\hline 
 $\Delta m^2_A$ & $|m_3^2-m_2^2|$ & $2.40\times 10^{-3}~{\rm eV^2}$ & $0.12\times 10^{-3}~{\rm eV^2}$\tabularnewline
\hline 
 $\sin\theta_S$ & $\sin\theta_{12}$ & $0.551$ & $0.017$ \tabularnewline
\hline 
 $\sin\theta_A$ & $\sin\theta_{23}$ & $0.707$ & $0.046$\tabularnewline
\hline 
 $\sin\theta_R$ & $\sin\theta_{13}$ & $0.1$ & $<0.14$ \tabularnewline
\hline 
 $m_{\nu_e}$ & $\sqrt{m_1^2c_{12}^2c_{13}^2 + m_2^2s_{12}^2c_{13}^2 + m_3^2s_{13}^2}$ & 0 & $0.50~{\rm eV}$\tabularnewline
\hline 
 $\Sigma$ & $m_1 + m_2 + m_3$  & 0 & $0.24~{\rm eV}$\tabularnewline
\hline
\hline 
 $m_{ee}$ & $|m_1c_{12}^2c_{13}^2 + m_2s_{12}^2c_{13}^2e^{2i\phi_2} + m_3s_{13}^2e^{2i(\phi_3-\delta)}|$  & 0 & $<0.175$\tabularnewline
\hline
\end{tabular}
\caption{Summary table of current neutrino results together with naming conventions and parameter definitions.  Columns three and four list best fit central parameter values and $1\sigma$ uncertainties.  The first five entries were adapted from the global oscillation analysis of \cite{ThreeFlavorOscUpdate}. The upper bounds for the last three entries come from endpoint measurements of Tritium $\beta$-decay ($m_{\nu_e}$), cosmological observations ($\Sigma$) and $\beta\beta0\nu$, respectively \cite{ImplicationsOfNuDataCirca2005}.  These are used as model constraints through the remainder of this analysis.}
\label{tab:Constraints}
\end{table}

The entries of the symmetric matrix $m_{\alpha\beta}$ are constrained to be small by a combination of the kinematic and oscillation neutrino data.  The suppression of this term with respect to the other charged fermions is likely due to the high scale of new physics.  The mass term of Eq.~(\ref{eq:MajMassLag}) is the relic of an effective operator after electroweak symmetry breaking and as such, $m_{\alpha\beta} \propto v\left(\frac{v}{\Lambda}\right)^n a_{\alpha\beta}$ where $v=0.246~{\rm TeV}$ is the Higgs vev and $a_{\alpha\beta}$ is a matrix of complex constants.  This may arise from a simple dimension five operator as in the seesaw mechanisms \cite{SeeSaw3Paths,SSF1,SSF2,SSF3,SSF4,SSF5} or from more exotic high dimensional interactions \cite{RadIndMajMassZee,MyLNV,LNVEffOpClassBabu}.  The specific UV completion is irrelevant for the purposes of this paper.  All that is required is a possibly broken flavor symmetry principle that is manifest in the low energy effective system.  Then the neutrino fields of Eq.~(\ref{eq:MajMassLag}) will transform under a representation of the symmetry group and only those mass matrices $m_{\alpha\beta}$ that render $\mathcal{L}_\nu$ invariant will be allowed.  Some of these symmetries will require particular texture zeros for full invariance that when broken will induce small deviations from zero that may be probed experimentally.

Assuming the Majorana nature of neutrinos, I extract the minimal value
of $m_{ee}$ for a variety of model classes.  Abelian flavor symmetries are useful in this endeavor due to the freedom of charge assignments to the neutrino fields.  Under non-Abelian symmetries, on the other hand, the fields will transform under some representation of the group and the Lagrangian terms will be composed of invariant field combinations.  Members of the field multiplets are assigned quantum numbers, as in the Abelian case, except that these are imposed by the group representation.  These charges are an important factor in mass matrix construction and may be mimicked by a properly constructed Abelian symmetry.  There is more freedom in mass matrix construction associated with Abelian symmetries, which proves to be useful when scanning for small $m_{ee}$ values.  The restrictions imposed by non-Abelian symmetries can only push the extracted $m_{ee}$ values up.  Thus, in what follows, the discussion is restricted to Abelian flavor symmetries.

\subsection{Unbroken Abelian Flavor Symmetries \label{sub:UnbrokenAbelian}}

Here I assume that a zero $m_{ee}$ element is protected by an Abelian
flavor symmetry. Given the $U(1)_{f}$ charges $n_{e}$, $n_{\mu}$
and $n_{\tau}$ supplied respectively to the $\nu_{e}$, $\nu_{\mu}$
and $\nu_{\tau}$ neutrinos\footnote{The neutrinos are components of the SM left-handed doublet fields $L_\alpha = \left(\begin{array}{c}
\nu_\alpha \\
\ell_{\alpha}\end{array}\right)$, thus left-handed charged leptons $\ell_{\alpha}$ will also be charged under $U(1)_f$.  The symmetry dictates the identity of the flavor basis by well defined charge assignments.  For my purposes, freedom in the flavor charge structure of the right-handed leptons $e_R$ can be used to propertly construct the diagonal charged lepton mass matrix. }, it is simple to derive the form of the
resulting symmetric Majorana neutrino mass matrix invariant under $U(1)_f$
\begin{equation}
m_{\alpha\beta}=Ma_{\alpha\beta}\delta_{(n_{\alpha}+n_{\beta}),0}.\label{eq:AblSymMassMatrix}
\end{equation}
$a_{\alpha\beta}$ is some, presumably $\mathcal{O}(1)$, complex
constant and $M\sim v\left(\frac{v}{\Lambda}\right)^n$ is the neutrino mass scale.  Currently, $M$ is bounded at
$0.05$ eV from below by the atmospheric mass squared difference \citep{ImplicationsOfNuDataCirca2005,StatusGlobFitsMaltoni07}
and from above at roughly $1$ eV by cosmological data \citep{ImplicationsOfNuDataCirca2005}. Notice that the electron
neutrino mass term $m_{ee}\overline{\nu_{e}^{c}}\nu_{e}$ has charge
$2n_{e}$. To preserve the imposed flavor symmetry, either $n_{e}=0$
or $m_{ee}=0$. Thus, a non-trivial transformation of $\nu_{e}$ under
$U(1)_{f}$ guarantees $m_{ee}=0$ and consequently a vanishing $\beta\beta0\nu$
rate. If one assumes that all allowed entries are nonzero, there are
eleven possible $m_{ee}=0$ mass matrix classes that can be obtained
from Eq.~(\ref{eq:AblSymMassMatrix}) by scanning charge assignments\footnote{The trivially zero mass matrix, obtained when $n_{\alpha}+n_{\beta}\neq0$
for all flavors $\alpha$ and $\beta$, is not included in this listing.%
}. These are listed in Table \ref{tab:AblSummary}, using the charge
assignment notation $(n_{e},n_{\mu},n_{\tau})$, along with their
associated mass matrix and neutrino mixing predictions.
There are only a small number, between one and three, of free parameters in each matrix entry.  These must be used to construct three mixing angles and two mass squared difference
predictions upon diagonalization. Thus, one obtains large correlations among the derived
oscillation parameters. If the coupling coefficients $a_{\alpha\beta}$
are allowed to take on any value, the class assignments are superfluous
in that some entries are just special limiting cases of other classes.
For example, $C11$ is just a special case of $C1$ with $a_{\mu\mu}=a_{\tau\tau}=0$.
These distinctions are made here due to the expectation
that all matrix elements allowed by $U(1)_f$ should be of the same order, in which case each
class yields different predictions. The predictions are obtained by
a simple diagonalization of the resulting mass matrix under the convention
that the smallest mass squared difference defines the solar oscillation frequency and
the next largest the atmospheric. The largest mass squared difference is the sum
of the smaller two and converges to $\Delta m^2_A$ when $\Delta m^2_S/\Delta m^2_A$
is small, as required by data. Degenerate eigenvalues are treated as if they possessed
small splittings induced by symmetry breaking effects in anticipation of Subsection \ref{sub:Broken-Abelian-Flavor}. The split levels are then associated with the
solar mixing sector and interpreted as such to make mass hierarchy predictions.
Even then, the neutrino mass ordering can only be predicted when the
lightest eigenvalue vanishes.  This occurs in all classes except $C1$,
$C3$ and $C6$. In these cases, one may derive relationships between
the mixing parameters and discrete hierarchy choices. Degeneracies
leading to invariant matrix subspaces in classes $C2$ and $C4$ yield
additional freedom corresponding to an arbitrary rotation within the
invariant subbasis. This is parameterized by a mixing angle $\theta$
in Table \ref{tab:AblSummary} that may take on any value. It should be noted that symmetry
breaking effects of $U(1)_{f}$, which select a definite mass basis, destroy
this freedom by selecting a particular value of $\theta$. Here, the symmetry breaking mechanism yields discrete variable changes and is therefore more important than in the other cases where deviations from the predictions
of Table \ref{tab:AblSummary} are parametrically small, or proportional
to the symmetry breaking order parameter. 
The goal of this exercise is to understand how close each $U(1)$
symmetry structure comes to reproducing the current neutrino oscillation
data, and in what ways they tend to fail. 
\begin{table}
\begin{tabular}{|c|c|c|>{\centering}m{3.5in}|}
\hline 
Class & Charge & Matrix & Predictions\tabularnewline
\hline
\hline 
$C1$ & $(n_{e},0,0)$ & $M\left(\begin{array}{ccc}
0 & 0 & 0\\
0 & a_{\mu\mu} & a_{\mu\tau}\\
0 & a_{\mu\tau} & a_{\tau\tau}\end{array}\right)$ & $s_{R}=0$, $t_{S}=0$ (Normal Hierarchy)

$t_{A}=0$, $t_{S}=0$ (Inverted Hierarchy)

Can tune $\Delta^{2}m$'s to fit data. See text for details.\tabularnewline
\hline 
$C2$ & $(n_{e},0,n_{\tau})$ & $M\left(\begin{array}{ccc}
0 & 0 & 0\\
0 & a_{\mu\mu} & 0\\
0 & 0 & 0\end{array}\right)$ & $\Delta m^2_S=0$, \textcolor{black}{$\Delta m^2_A=M^2a_{\mu\mu}^{2}$}

$s_{R}=\sin\theta$, $t_{A}=\cos\theta$, $t_{S}=0$

Normal Hierarchy \tabularnewline
\hline 
$C3$ & $(n_{e},0,-n_{e})$ & $M\left(\begin{array}{ccc}
0 & 0 & a_{e\tau}\\
0 & a_{\mu\mu} & 0\\
a_{e\tau} & 0 & 0\end{array}\right)$ & $\Delta m^2_S=0$, \textcolor{black}{$|\Delta m^2_A|=M^2|a_{\tau\tau}^{2}-a_{e\mu}^{2}|$}

$s_{R}^{2}=1/2$, $t_{A}=0$, $t_{S}=0$

Any Hierarchy\tabularnewline
\hline 
$C4$ & $(n_{e},n_{\mu,}0)$ & $M\left(\begin{array}{ccc}
0 & 0 & 0\\
0 & 0 & 0\\
0 & 0 & a_{\tau\tau}\end{array}\right)$ & $\Delta m^2_S=0$, \textcolor{black}{$\Delta m^2_A=M^2a_{\tau\tau}^{2}$}

$s_{R}=0$, $t_{A}=0$, $t_{S}=\tan\theta$

Normal Hierarchy\tabularnewline
\hline 
$C5$ & $(n_{e},n_{e},-n_{e})$ & $M\left(\begin{array}{ccc}
0 & 0 & a_{e\tau}\\
0 & 0 & a_{\mu\tau}\\
a_{e\tau} & a_{\mu\tau} & 0\end{array}\right)$ & $\Delta m^2_S=0$, \textcolor{black}{$\Delta m^2_A=M^2(a_{e\tau}^{2}+a_{\mu\tau}^{2})$}

$s_{R}^{2}=1/2$, $t_{A}=0$, $t_{S}=a_{\mu\tau}/a_{e\tau}$

Inverted Hierarchy\tabularnewline
\hline 
$C6$ & $(n_{e},-n_{e},0)$ & $M\left(\begin{array}{ccc}
0 & a_{e\mu} & 0\\
a_{e\mu} & 0 & 0\\
0 & 0 & a_{\tau\tau}\end{array}\right)$ & $\Delta m^2_S=0$,\textcolor{blue}{ }\textcolor{black}{$|\Delta m^2_A|=M^2|a_{\tau\tau}^{2}-a_{e\mu}^{2}|$}

${\color{blue}{\color{black}s_{R}=0}}$, $t_{A}=0$, $t_{S}=1$

Any Hierarchy\tabularnewline
\hline 
$C7$ & $(n_{e},-n_{e},n_{e})$ & $M\left(\begin{array}{ccc}
0 & a_{e\mu} & 0\\
a_{e\mu} & 0 & a_{\mu\tau}\\
0 & a_{\mu\tau} & 0\end{array}\right)$ & $\Delta m^2_S=0$, \textcolor{black}{$\Delta m^2_A=M^2(a_{\mu\tau}^{2}+a_{e\mu}^{2})$}

$s_{R}^{2}=\frac{a_{\mu\tau}^{2}}{2(a_{e\mu}^{2}+a_{\mu\tau}^{2})}$,
$t_{A}^{2}=\frac{a_{\mu\tau}^{2}}{2a_{e\mu}^{2}}$, $t_{S}^{2}=\frac{a_{e\mu}^{2}+a_{\mu\tau}^{2}}{a_{e\mu^{2}}}$

Inverted Hierarchy, $s_{R}=t_{A}/t_{S}$\tabularnewline
\hline 
$C8$ & $(n_{e},-n_{e},-n_{e})$ & $M\left(\begin{array}{ccc}
0 & a_{e\mu} & a_{e\tau}\\
a_{e\mu} & 0 & 0\\
a_{e\tau} & 0 & 0\end{array}\right)$ & $\Delta m^2_S=0$, \textcolor{black}{$\Delta m^2_A=M^2(a_{e\tau}^{2}+a_{e\mu}^{2})$}

$s_{R}^{2}=\frac{a_{e\tau}^{2}}{2(a_{e\mu}^{2}+a_{e\tau}^{2})}$,
$t_{A}^{2}=\frac{a_{e\tau}^{2}}{2a_{e\mu}^{2}}$, $t_{S}^{2}=\frac{a_{e\mu}^{2}}{2(a_{e\mu}^{2}+a_{e\tau}^{2})}$

Inverted Hierarchy, $s_{R}=t_{A}t_{S}$\tabularnewline
\hline 
$C9$ & $(n_{e},-n_{e},n_{\tau})$ & $M\left(\begin{array}{ccc}
0 & a_{e\mu} & 0\\
a_{e\mu} & 0 & 0\\
0 & 0 & 0\end{array}\right)$ & $\Delta m^2_S=0$, \textcolor{black}{$\Delta m^2_A=M^2a_{e\mu}^{2}$}

$s_{R}=0$, $t_{A}=0$, $t_{S}=1$

Inverted Hierarchy\tabularnewline
\hline 
$C10$ & $(n_{e},n_{\mu},-n_{e})$ & $M\left(\begin{array}{ccc}
0 & 0 & a_{e\tau}\\
0 & 0 & 0\\
a_{e\tau} & 0 & 0\end{array}\right)$ & $\Delta m^2_S=0$, \textcolor{black}{$\Delta m^2_A=M^2a_{e\tau}^{2}$}

$s_{R}^{2}=1/2$, $t_{A}=0$, $t_{S}=0$

Inverted Hierarchy\tabularnewline
\hline 
$C11$ & $(n_{e},n_{\mu},-n_{\mu})$ & $M\left(\begin{array}{ccc}
0 & 0 & 0\\
0 & 0 & a_{\mu\tau}\\
0 & a_{\mu\tau} & 0\end{array}\right)$ & $\Delta m^2_S=0$, \textcolor{black}{$\Delta m^2_A=M^2a_{\mu\tau}^{2}$}

$s_{R}^{2}=1/2$, $t_{A}=0$, $t_{S}=0$

Inverted Hierarchy\tabularnewline
\hline
\end{tabular}

\caption{Exhaustive summary of neutrino mass matrices with $m_{ee}=0$ protected by Abelian flavor symmetries.  Column one defines the symmetry class as referred to throughout the text.  Columns two and three list representative $U(1)_f$ charge assignments to the neutrino flavor eigenstates and the implied mass matrix, respectively.  The final column shows oscillation parameter predictions for each class.  None of these are consistent with current neutrino data.  See the text for more information.}

\label{tab:AblSummary}
\end{table}

Most predictions of Table \ref{tab:AblSummary} are well defined
and need little explanation. $C1$, however, is more involved and
deserves a separate discussion for clarity. Changing the parametrization
of the mass matrix for convenience to
\begin{equation}
\left(\begin{array}{ccc}
0 & 0 & 0\\
0 & a-b & \pm\sqrt{d^{2}-b^{2}}\\
0 & \pm\sqrt{d^{2}-b^{2}} & a+b\end{array}\right),
\end{equation}
I find that the lightest mass eigenvalue is zero, while the absolute
mass squared differences are $(a-d)^{2}$ and $4ad$. The associated atmospheric and solar parameters depend on their relative splitting sizes.
We are left with two cases defined by their predicted mass hierarchy.
For the inverted hierarchy, $\Delta m^2_S=(a-d)^{2}$ and $\Delta m^2_A=4ad$,
while all mixing angles vanish except $\theta_{R}$, which is given by
$s_{R}^{2}=(d-b)/2d$. For the normal hierarchy, the mass squared
differences are reversed and all mixing angles vanish except $\theta_{A}$,
which is given by $t_{A}^{2}=(d-b)/(d+b)$. It is interesting that
this is the only case, due to availability of three free mass matrix parameters,
that allows for a nonzero solar mass squared difference.  Consequently, this class is well-suited to fit the neutrino data with only minor modifications.  While neither
case is consistant with neutrino oscillation phenomenology, it is clear that the normal hierarchy choice
can be pushed ``closer'' to the observed form. For nearly maximal atmospheric
mixing, $b$ must be small. For the mass squared differences to work
out, the parameters $a$ and $d$ must be unnaturally tuned to $a/4d\approx10^{\pm2}$.
Hence, this scenario is far from ideal when considered with universally
$\mathcal{O}(1)$ parameter values.  This mass matrix texture is theoretically motivated by variants of $\mu-\tau$ symmetries and is commonly found in the literature.  See for example \citep{MyLNV,StructureNuMassMatrixCpViolation,ElementsOfNuMassMatrix_AllowedRangesTextureZeros}
and references therein. 

Each entry of Table \ref{tab:AblSummary} defines a class of models
with similar characteristic predictions. It is important to note that
none of these classes fit the neutrino mixing data. This is another
reiteration of the fact that for Majorana neutrinos, $m_{ee}\neq0$,
which implies a nonzero $\beta\beta0\nu$ rate. In particular, none
of the classes predict realistic mixing angles. This can be seen by
inspection, as most cases predict some combination of maximal $\theta_{R}$
or vanishing solar/atmospheric angles. Classes $C7$ and $C8$ are less trivial
but still ultimately fail. $C8$ fails due to the relationship between
the angles $s_{R}=t_{S}t_{A}$, implying that a small $\theta_{R}$
must be accompanied by either a small $\theta_{S}$ or $\theta_{A}$.
Similarly, in $C7$ $\theta_{S}$ must be close to $\pi/2$ to
insure a small reactor angle. Additionally, all classes except the
first yield degenerate eigenvalues implying $\Delta m^2_S=0$, which
further contradicts observation.

\subsection{Broken Abelian Flavor Symmetries \label{sub:Broken-Abelian-Flavor}}

Exact $m_{ee}=0$ may be excluded by this line of reasoning, but it
still may be unobservably small. In this case, the symmetries of Table
\ref{tab:AblSummary} should still be approximately valid, only broken
by some small amount $\epsilon$. These broken scenarios should still
retain some of the features of their parent classes, such as mass hierarchy
or large/small mixing angles. Hence, one would expect that classes
such as $C1$, $C2$, and $C4$ with multiple predictions that can be made consistant with data, will be broken
far less than, say, classes $C10$ and $C11$ that are far from data.

I parameterize this symmetry breaking with the introduction of a spurious
scalar field $s$, charged under $U(1)_{f}$, that acquires a nonzero
vev. For this spurion analysis, $s$ is just
a mathematical construct used to understand the pattern and size of
symmetry breaking. Specifically, I assume
that $s$ has $U(1)_f$ charge $n_{s}$ and acquires a real vev $\epsilon$.
With this, small mass term corrections are induced to help fit the data. The
form of the resulting mass matrix may now be described by a 4-tuple
$(n_{e},n_{\mu},n_{\tau};n_{s})$. For example, the symmetry of class
$C6$ may be broken to yield
\begin{equation}
m(n_{e},-n_{e},0;n_{e})=M\left(\begin{array}{ccc}
a_{ee}\epsilon^{2} & a_{e\mu} & a_{e\tau}\epsilon\\
a_{e\mu} & a_{\mu\mu}\epsilon^{2} & a_{\mu\tau}\epsilon\\
a_{e\tau}\epsilon & a_{\mu\tau}\epsilon & a_{\tau\tau}\end{array}\right),
\end{equation}
or 
\begin{equation}
m(n_{e},-n_{e},0;2n_{e})=M\left(\begin{array}{ccc}
a_{ee}\epsilon & a_{e\mu} & 0\\
a_{e\mu} & a_{\mu\mu}\epsilon & 0\\
0 & 0 & a_{\tau\tau}\end{array}\right),
\end{equation}
depending on the spurion charge assignment.  In what follows, I will refer to these broken structures by integer valued 4-tuples.  In the previous example, this would be $(1,-1,0;1)$ and $(1,-1,0;2)$, respectively.  These are not unique but adequately represent a broken class.  In parent classes $C1$, $C3$, and $C5-C8$ with a single assigned charge, the non-trivial\footnote{Nontrivial in this context refers to an assignment that yields perturbed mass matrices distinct from the parent class structure.} spurion charges are always multiples of $n_e$.  These cases yield between two and three perturbed mass matrix structures for each class.  For the remainder of the parent classes, with two distinct charge values, non-trivial spurion charges must be multiples of either neutrino charge value or their sum.  This freedom leads to a proliferation of perturbed mass matrices - between 37 and 52 for each class.  For a general charge assignment, the lowest order mass matrix elements are given by 
\begin{equation}
m_{\alpha\beta}=M\sum_{i=0}^\infty a_{\alpha\beta}^{(i)} \epsilon^i\delta_{(n_\alpha + n_\beta),in_s}\label{eq:MabBrokenU1},
\end{equation}
where $a_{\alpha\beta}^{(i)}$ are complex constants for the $\mathcal{O}(\epsilon^i)$ terms.  Once again, it is reasonable to assume that all nonzero $a_{\alpha\beta}^{(i)}$ are of the same order of magnitude, since they all arise as coupling constants within the spurion included invariant Lagrangian.  Clearly, the delta function vanishes for all but a single $i$ value.  From here, it is trivial to build up the matrix to higher order.  If the operator $\mathcal{O}_i$ is some $U(1)_f$ invariant function of the neutrino and spurion fields, of order $\mathcal{O}(\epsilon^i)$ after flavor symmetry breaking, then the operator $\mathcal{O}_{i+2} = \overline{s}s\mathcal{O}_i$ is also invariant of order $\mathcal{O}(\epsilon^{i+2})$ after $s$ acquires a vev.  This process may be continued indefinitely to yield the full matrix.  Thus, each nonzero matrix element is expressable as a power series in $\epsilon^2$.  For most cases of interest, where $\epsilon$ is sufficiently small, all but the leading terms
 described by Eq.~(\ref{eq:MabBrokenU1}) may be neglected.  This new broken matrix may be diagonalized perturbatively to yield predictions for oscillation and kinematic parameters as a function of $\epsilon$.  Due to the number of charge assignments and breaking patterns, it is most efficient to study this numerically.

\subsubsection{Numerical Results\label{subsub:Numerical-Results} }

I now numerically survey all mass matrices generated from a broken
$U(1)$ flavor symmetry to order $\epsilon^{10}$ in $m_{ee}$ and $\epsilon^{3}$ in all other elements.  The unequal treatment of the mass matrix is due to the search goal of small $m_{ee}$.  Corrections higher than $\epsilon^3$ make little difference to the fit, but one must be able to distinguish cases with different large $m_{ee}$ suppressions, even when the rest of the matrix is identical.  It will turn out that no structure with $m_{ee}$ suppressed by more then $\epsilon^8$ is consistant with data.  Thus, truncating the search at $\epsilon^{10}$ is safe. Using the notation
of Eq.~(\ref{eq:MabBrokenU1}), I set $a_{ee}=1$ in order to remove
the ambiguity of simultaneous rescalings of all the $a_{\alpha\beta}$
and $M$. Furthermore, I rescale $\epsilon^{\prime}\rightarrow\epsilon^{\gcd(P)}$,
where $P$ stands for the set of all nonzero powers of $\epsilon$
in the original mass matrix and $\gcd$ is the greatest common denominator function. This removes another unphysical ambiguity.  To see this, suppose that two different charge assignments yield the same perturbed matrix structure up to an overall rescaling of $\epsilon \rightarrow \epsilon^j$.  In this case, the best fit to the data would select out preferred $\epsilon$ values related by the rescaling, but the same mass matrix elements.  Thus, from the neutrino mass generation standpoint, both charge assignments yield the same predictions and one should be selected to represent the system\footnote{This is not true when a spurion analysis beyond neutrino mass is performed.  In that case, each assignment would yield distinct predictions for other processes.}.
After all rescalings and truncations, there are 230 distinct possibilities that may be categorized into
one of the eleven classes of Table \ref{tab:AblSummary}. In each
case, I scan the symmetry breaking parameter $\epsilon$, the mass scale $M$ and coupling constants $a_{\alpha\beta}^{(i)}$ to fit current
neutrino constraints listed in Table \ref{tab:Constraints} and extract the smallest allowed $m_{ee}$. 

Specifically, for each model I perform a $\chi^2$ fit to minimize the function
\begin{equation}
\chi^2(M,a_{\alpha\beta}^{(i)}) = \sum_{j = \{{\rm S,A}\}} \frac{(\Delta^2m_j^\prime-\Delta^2m_j)^2}{\sigma^2_{\Delta^2m_j}} + \sum_{j = \{{\rm S,A,R}\}} \frac{(\sin\theta_j^\prime-\sin\theta_j)^2}{\sigma^2_{\sin\theta_j}} + \frac{(m_\nu^\prime)^2}{\sigma^2_{m_\nu}} + \frac{(\Sigma^\prime)^2}{\sigma^2_\Sigma} \label{eq:Chi2},
\end{equation}
where the primed quantities are evaluated from the broken mass matrix and the unprimed best fit values and uncertainties are taken from Table \ref{tab:Constraints}. Here, only $M$ and $a_{\alpha\beta}^{(i)}$ are varied to yield the one parameter function $F(\epsilon)$.  When $F(\epsilon)$ falls below a critical value, the broken flavor symmetry is allowed at a specified confidence.  To be conservative, I use $99\%$ confidence limits throughout this analysis.  The critical $\chi^2$ value depends on the number of degrees of freedom within the system and is therefore different for each case.  The allowed domain where $F(\epsilon)$ falls below its critical value may then easily be scanned for the smallest $m_{ee}$ value.  If no such region exists, the charge assignment and breaking structure is disfavored by current data at $99\%$ confidence. 
To maintain the perturbativity of the system, I hold $\epsilon<2/3$;
in which case, the largest possible (fourth order) corrections are
only $(2/3)^{4}\approx20\%$. Typically, corrections will be much
smaller than this since $m_{ee}$ goes like some power of $\epsilon$
and small $m_{ee}$ values favor small $\epsilon$. For this procedure
to make sense, the constants $|a_{\alpha\beta}|$ must be constrained
by some naturalness criterion, else the small $\epsilon$ values may
be compensated by large coupling constants resulting in the loss of algebraic
structure information. In this spirit, I only allow the constants
to vary symmetrically about unity, in a $\log_{10}$ sense, by a
small amount. That is, $10^{-c}<|a_{\alpha\beta}|<10^{c}$ for some small
number $c$. A one and two order of magnitude spread is defined
by $c=0.5$ and $c=1$, respectively. Ideally, the relative size of
each matrix element should be determined by $\epsilon$ alone, so it
is clear that $c$ should not be much greater than $0.5$ for a typical
$\epsilon\approx10^{-1}$. Even this range is dangerous near the
upper $\epsilon$ limit where order $\epsilon^{n}$ terms can easily
be larger than order $\epsilon^{n-1}$ terms. To remove this problem
completely for the full $\epsilon$ range, one needs $c\lesssim0.1$
which offers very little parameter freedom and is not realistic. I present data for $c=0.1$, $0.3$, $0.5,$
and $0.7$ in the attempt to cover a comprehensive range of naturalness
criterion.

\begin{figure}
\includegraphics[bb=25bp 315bp 638bp 851bp,clip,scale=0.790]{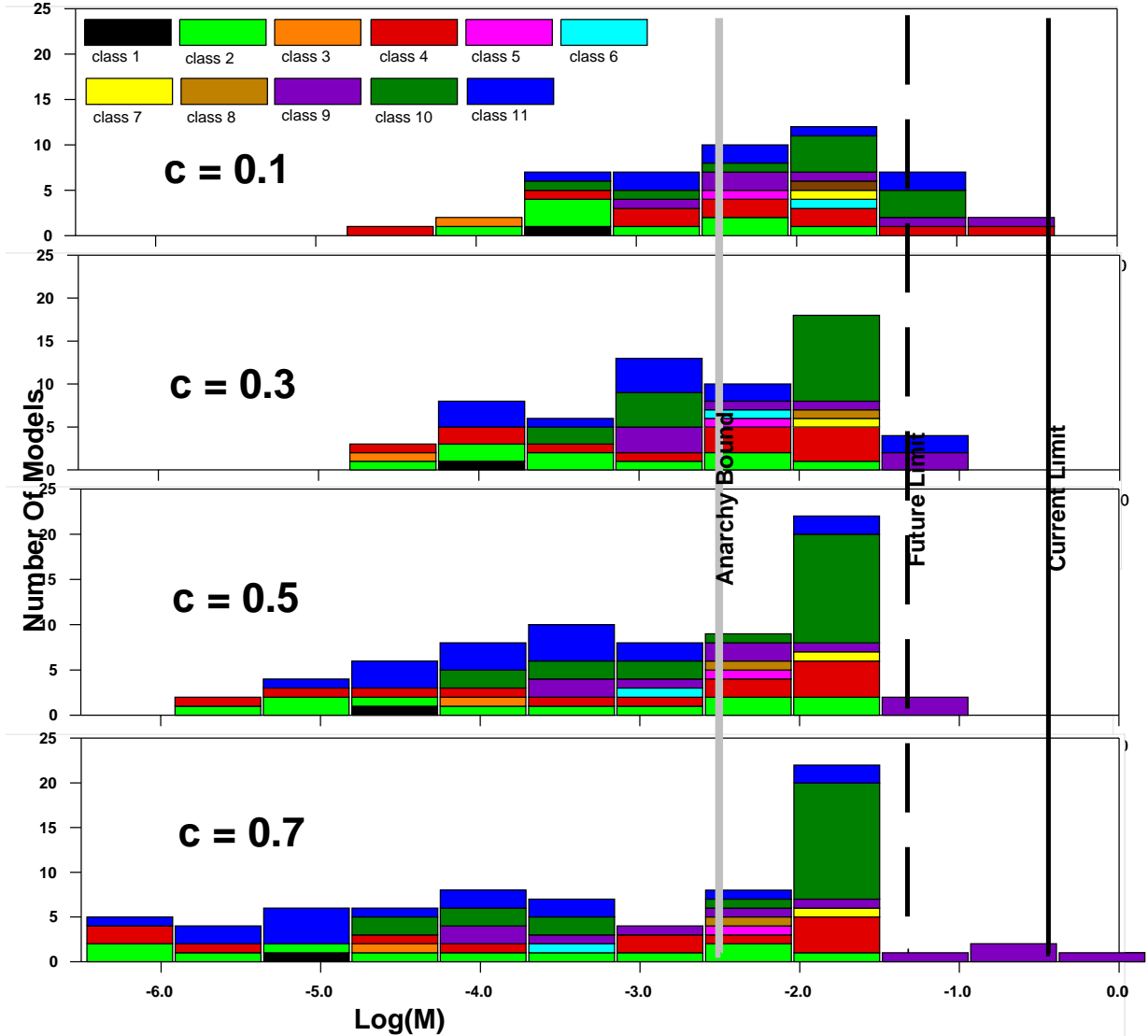}
\caption{Histogram of minimal $m_{ee}$ values extracted from data allowed models, color coded by model class.  From top to bottom, the panels represent the $c=0.1$, $c=0.3$, $c=0.5$ and $c=0.7$ cases.  $c$ defines fine tuning sensitivity by limiting the relative magnitudes of the free matrix parameters as $10^{-c}<|a_{\alpha\beta}|<10^{c}$.  A larger $c$ value represents more parameter freedom, which leads to a greater number of allowed models and smaller $m_{ee}$, as is clear from the plot.  See the text for more details.}\label{fig:Total4Hist}
\end{figure}

Figure \ref{fig:Total4Hist} histograms the $99\%$ allowed flavor structures by minimum $m_{ee}$ value for $c=0.1$, $0.3$, $0.5$ and $0.7$ on a $\log_{10}$ scale.  These are color-coded by parent class.  Reference lines indicating the $\beta\beta0\nu$ current bound (solid black) and future reach (broken black) are included for reference as well as the anarchy bound (solid gray).  Due to increasing parameter freedom, the number of allowed models increases and broadens as the $c$ value is pushed higher.  For example, the $57$ models of $c=0.1$ span less than four orders of magnitude, while the $74$ models of $c=0.7$ span seven. The smallest $m_{ee}$ values for $c=0.1$, $0.3$, $0.5$ and $0.7$ are $6.43\times 10^{-5}~{\rm eV}$, $2.96\times 10^{-5}~{\rm eV}$, $3.92\times 10^{-6}~{\rm eV}$ and $5.28\times 10^{-7}~{\rm eV}$, respectively.  Optimally minimized charge assignments along with their relevant parameters are summarized in Table \ref{tab:MinModels}.  The mean $\epsilon$ values for each $c$ is also shown for easy comparison.  The broken symmetry that yields the smallest rates are $(4,-1,0;1)$ for the $c=0.1$, $c=0.3$ and $c=0.5$ cases and $(4,0,-1;1)$ for the $c=0.7$ case.  The last symmetry, optimal for $c=0.7$, only beats out $(4,-1,0;1)$ by $0.73\%$.  These are members of very similar parent classes $C4$ and $C2$.  It is not a coincidence that these both contain invariant subspaces that allow for additional parameter freedom and predict the normal mass hierarchy.  $m_{ee}$ is so small in these cases because it is suppressed by $\epsilon^8$.  The reason there is so great a difference between them is due solely to the increased parameter freedom of higher $c$ values allowing for smaller $\epsilon$.  Even these tiny variations in $\epsilon$ are amplified in the $m_{ee}$ relations and can easily yield order of magnitude differences.

\begin{table}
\begin{tabular}{|c|c c c c c c c c c|}
\hline 
$c$ Case & Class & Charge & $m_{ee}~~({\rm eV})$ & $\epsilon$ & $<\epsilon>$& $\sin\theta_R$ & $\sin\theta_A$ & $\sin\theta_S$ & Hierarchy \\
\hline
\hline
$0.1$ & $C4$ & $(4,-1,0;1)$ & $6.43\times 10^{-5}$ & $0.402$ & $0.398$ & $0.0870$ & $0.5648$ & $0.5341$ & N \\
\hline
$0.3$ & $C4$ & $(4,-1,0;1)$ & $2.96\times 10^{-5}$ & $0.376$ & $0.367$ & $0.0863$ & $0.5793$ & $0.5293$ & N \\
\hline
$0.5$ & $C4$ & $(4,-1,0;1)$ & $3.92\times 10^{-6}$ & $0.276$ & $0.279$ & $0.0859$ & $0.5945$ & $0.5217$ & N \\
\hline
$0.7$ & $C2$ & $(4,0,-1;1)$ & $5.38\times 10^{-7}$ & $0.203$ & $0.218$ & $0.0859$ & $0.5932$ & $0.5207$ & N \\
      & $C4$ & $(4,-1,0;1)$ & $5.32\times 10^{-7}$ & $0.203$ &         & $0.0862$ & $0.5966$ & $0.5200$ & N \\
\hline 
\end{tabular}
\caption{Summary of models with the smallest $m_{ee}$ values for the $c=0.1$, $c=0.3$, $c=0.5$ and $c=0.7$ cases, as listed in column one.  Columns two, three and four list the model's parent class, a representative charge assignment and derived $m_{ee}$.  Column five shows the optimal amount of $U(1)_f$ symmetry breaking via the order parameter $\epsilon$ to be compared with the mean value $<\epsilon>$ in column six.  The last four columns give mixing angle and mass hierarchy predictions.  Two items with similar $m_{ee}$ are listed for case $c=0.7$ for easy comparison.  See the text for details.}\label{tab:MinModels}
\end{table}

Inspection of the class descriptions of Table \ref{tab:AblSummary} reveals that both classes $C2$ and $C4$ have two distinct problems that must be solved by symmetry breaking.  Specifically, it must induce a nonzero $\Delta m^2_S$, as well as push $\theta_S$ and $\theta_A$, respectively, up to allowed levels.  Additionally, the symmetry breaking mechanism must be able to explain the $\theta$ values required by data, namely $\theta \sim 0$ and $\theta \sim \theta_S$ for classes $C2$ and $C4$, respectively.  Looking at the broken $C4$ case $(4,-1,0;1)$ as a representative example, it is easy to see how these are solved.  An allowed real valued matrix corresponding to the $c=0.3$ case is
\begin{equation}
M\left(\begin{array}{ccc}
\left(\epsilon^8\right) & -1.42\epsilon^3 & 0\\
-1.42\epsilon^3 & 2.00\epsilon^2 & 0.78\epsilon\\
0 & 0.78\epsilon & 0.50
\end{array}\right) + \mathcal{O}(\epsilon^4),
\end{equation}
where $M=0.074~{\rm eV}$ and $\epsilon = 0.37$.  The $m_{ee}\propto\epsilon^8$ is included in the matrix for illustrative purposes.  Some predictions of this structure are $\sin\theta_R \propto \epsilon^3$, $\sin\theta_S \propto \epsilon$ and $\Delta m^2_S/M^2 \propto \epsilon^4$ up to $\mathcal{O}(1)$ modifications. Taking the allowed $\Delta m^2_S$ and neutrino mass scale range in the last relation implies $\epsilon$ between $0.01$ and $0.25$ as observed in the fit.  The large solar mixing angle data selects the upper part of this range which leads to a large $\theta_R$ prediction.  Once the numerical factors are accounted for, this matrix structure is very similar to that of class $C1$, with the lower right $\mu-\tau$ block elements of the same order, disagreeing at most by $56\%$.  As previously discussed, $C1$ is a popular texture that is allowed, provided small symmetry breaking.  The charge assignment $(1,0,0,1)$ of class $C1$ fits the data in the $c=0.5$ scenario with $\epsilon = 0.0158$, yielding the $\epsilon^2$ suppressed $m_{ee}=1.97\times 10^{-5}~{\rm eV}$.  It turns out that the majority of the minimized matrices with $m_{ee}<10^{-4}~{\rm eV}$ acquire this form.  This is optimal from the small $m_{ee}$ perspective, since all data can be accommodated with a vanishingly small $m_{ee}$ element.  That is, the $e-e$ matrix element does not significantly contribute to the fit.  I find that no broken symmetry with an exact $e-e$ texture zero can yield this approximate $C1$ structure, but highly suppressed terms are possible.  For these cases, the size of $m_{ee}$ depends most critically on the $\epsilon$ power suppression as opposed to the amount of symmetry breaking.

A handful of these, particularly those of class $C9$, will be probed by next generation $\beta\beta0\nu$ experiments. For example, in the representative $c=0.5$ case, the class $C9$ charge assignment $(1,-1,-3,1)$ will be explored.  The breaking is optimized with $\epsilon = 0.38$, leading to an $\epsilon^2$ suppressed $m_{ee}=0.070~{\rm eV}$ with the inverted mass hierarchy.  It should be noted that null result bounds set by $\beta\beta0\nu$ experiments do constrain those models with higher minimal $m_{ee}$ values.  However, a measurement of $\beta\beta0\nu$ does not pick out a particular broken flavor model since all that is being plotted is the minimum $m_{ee}$ value.  In other words, all unconstrained models should be considered equal candidates.  Furthermore, bounds or measurements of small $m_{ee}$ only indicate a flavor symmetry when they fall below the anarchy bound near $5\times 10^{-3}~{\rm eV}$.  Above that value suppressed $\Gamma_{\beta\beta0\nu}$ could simply be a random fluctuation of an anarchical mass matrix.

\begin{figure}
\includegraphics[scale=0.395]{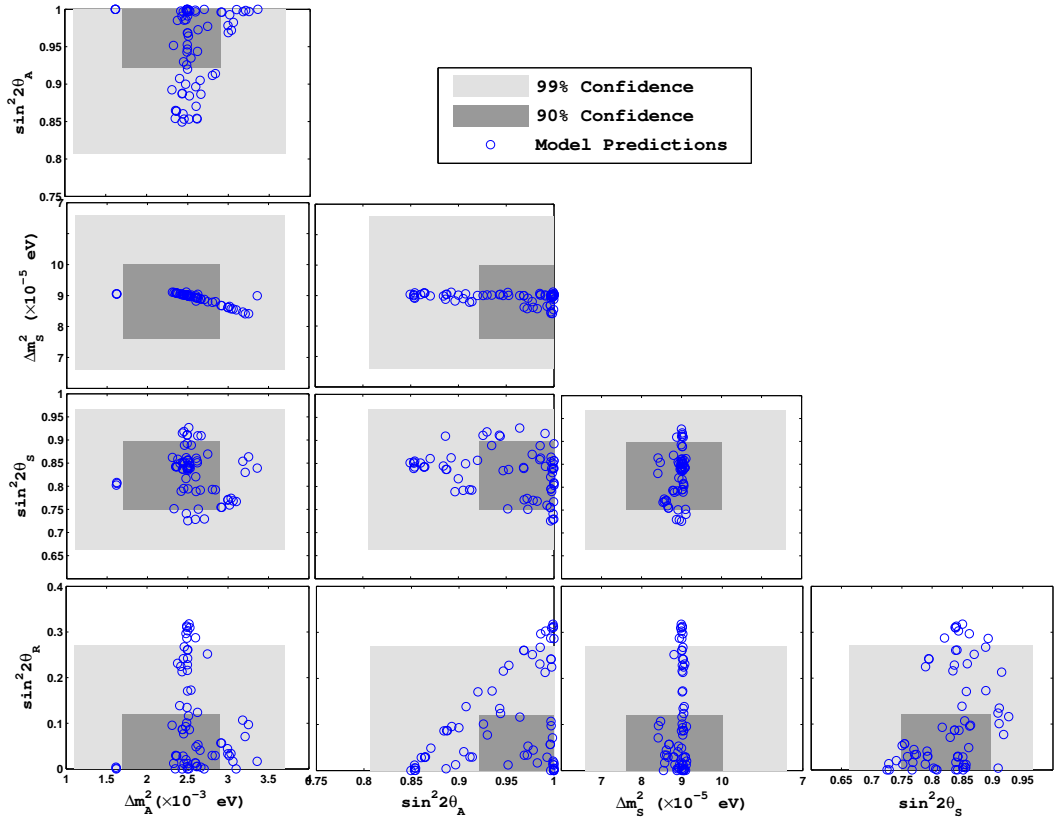}
\caption{Scatter plot projections of neutrino oscillation parameter $c=0.5$ predictions taken from data allowed models.  $90\%$ and $99\%$ parameter confidence limits are indicated by the dark and light gray regions, respectively.}
\label{fig:MixingScatter}
\end{figure}

Each of these models, taken at their minimum $m_{ee}$ values, make predictions for other observables.  These are found by diagonalizing the perturbed mass matrix, together with the minimization parameters, to obtain the mass eigenvalues and mixing angles.  They may also be combined, as prescribed in Table \ref{tab:Constraints}, to yield the mass squared differences as well as the effective neutrino mass relevant to tritium beta decay and the cosmological neutrino mass sum.  This prescription also selects the neutrino mass ordering predicted by each model.  Figure \ref{fig:MixingScatter} displays scatter plot predictions of the $c=0.5$ models allowed by data shown as projections onto the neutrino oscillation parameter space.  The dark and light gray shaded rectangles illustrate the $90\%$ and $99\%$ allowed regions, respectively.  These are not true confidence contours, as they do not take correlations into account, but adequately reflect parameter regions allowed by data for the purposes of this analysis.  Most of the scattered points are found to possess the normal mass ordering.  There are $64$ cases predicting normal and $7$ cases predicting inverted mass orderings.  Generally, the normal hierarchy cases yield a lower $m_{ee}$.  A quick inspection of this plot reveals that the mass squared differences, particularly $\Delta m^2_S$, have little spread and are thus not suited to distinguish between the symmetry structures. $\Delta m^2_A$ may be tuned to fit data in even the unbroken cases enumerated in Table \ref{tab:AblSummary} with minimal effects on remaining parameters.  Additionally, $\Delta m^2_S$ is a relatively small splitting that arises when $U(1)_f$ is broken in all classes but $C1$, where it is present from the beginning.  For models that allow such small splittings, it is easy to adjust parameters to fit the data.  Mixing angles, however, turn out to be a much better diagnostic tool.  The unbroken classes of Table \ref{tab:AblSummary} predict vanishing or maximal angles for the majority of entries.  If these fit the data as given, say, maximal $\theta_A$ and vanishing $\theta_{R}$, symmetry breaking is likely to drive these away from the preferred values.  In the opposite limit, when predictions are far from data, the symmetry breaking must drive the angle significantly to reach the allowed range.  This results in a large spread in predicted angles.  I find that $\theta_{R}$ and $\theta_A$ are the best laboratories for constraining the model classes.  The points in the projection onto this parameter plane spans almost the entire allowed region.  The triangular shape of the scatter profile in this panel is an artifact of the fitting procedure.  If a large deviation from best fit is found in one parameter, there is little room left for deviations in any other parameter.

\begin{figure}
\includegraphics[scale=.645]{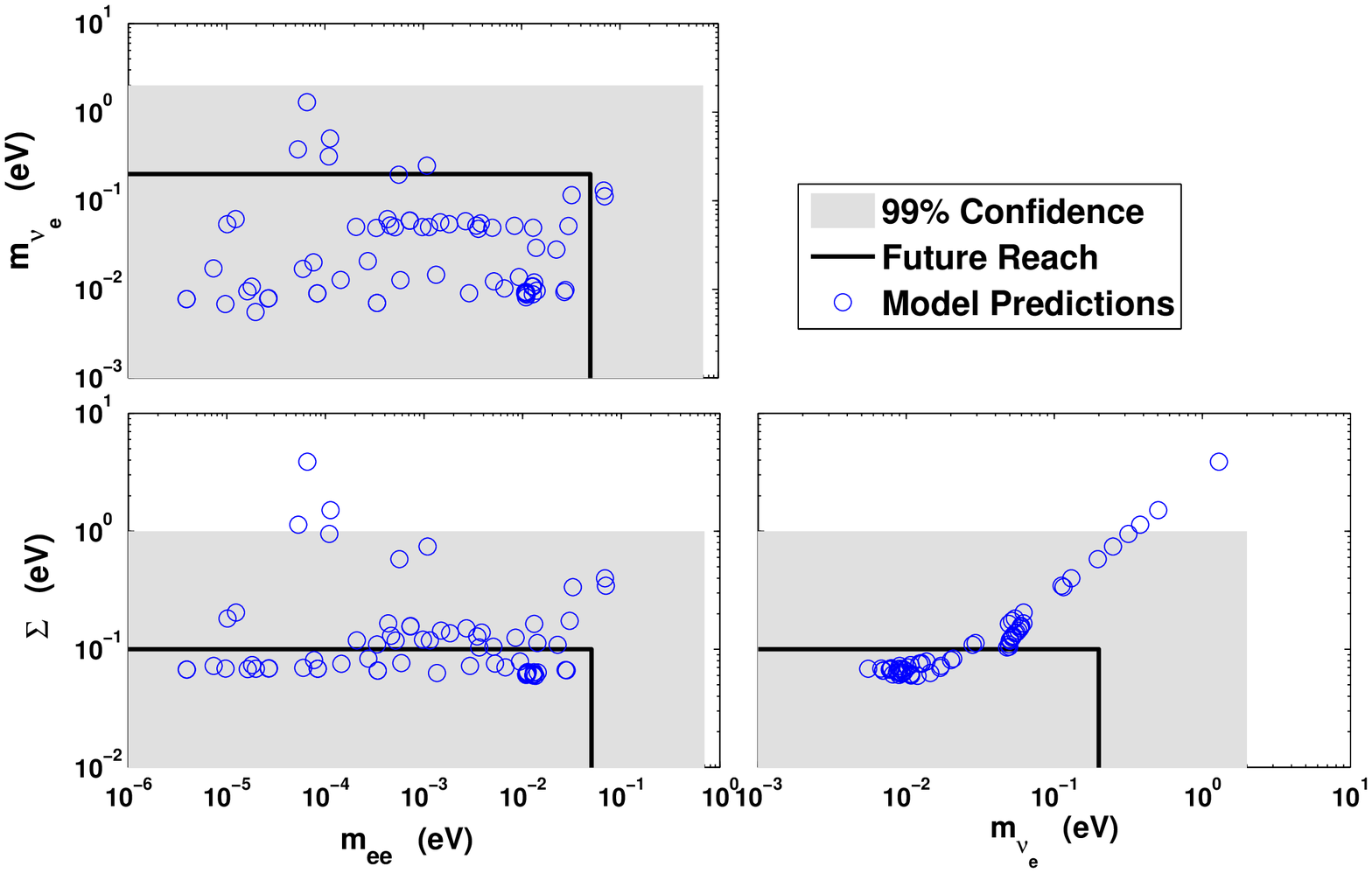}
\caption{Scatter plot projections of neutrino kinematic and absolute mass predictions ($m_{\nu_e}$, $\Sigma$ and $m{ee}$) taken from data allowed models.  Current $99\%$ confidence limits are indicated by the gray region, and the solid black contour line illustrates the next generation experimental reach.}
\label{fig:KinimaticScatter}
\end{figure}

In a similar way, Figure \ref{fig:KinimaticScatter} displays $c=0.5$ prediction scatter plots in the space of absolute mass observables $m_{\nu_e}$, $\Sigma$ and $m_{ee}$.  The $m_{ee}$ direction is histogrammed in the third panel of Figure \ref{fig:Total4Hist} and is included here to help visualize the relationship among the parameters.  Once again, light gray illustrates the $99\%$ allowed region.  The solid black contour line represents the reach of next generation experiments.  These contours only approximate the present and future experimental bounds, as there are large correlations among $m_{ee}$, $\Sigma$ and $m_{\nu_e}$ under the three light Majorana neutrino assumption employed here.  See for example \cite{AbsoluteNuScaleSpecBBon,MajNuBB0nCosmology,MajNuCPVBB0nTrit,MyNonOscHier,ObsAbsNuMassConstraintsCorr} and references therein for a discussion of these correlations. The smallest inverted hierarchy $m_{ee}=.0223~{\rm eV}$ is for the $C7$ charge structure $(1,-1,1;1)$.  This is the only class $C7$ model consistent with data, and thus requires a relatively large amount of symmetry breaking, with $\epsilon=0.3159$.  The remaining six inverted hierarchy cases are of classes $C9$, $C10$ and $C11$, consistent with expectations from the Table \ref{tab:AblSummary} predictions.  Within these, the hierarchy determination depends heavily on the specific charge structure.  Those predicting normal spectra generally have larger symmetry breaking in order to overcome the influence of their parent classes.  Nevertheless, these can have $m_{ee}$ as small as $10^{-5}~{\rm eV}$ due to large $\epsilon$ power suppressions.  Still, it is gratifying that the inverted cases are all above the $99\%$ confidence lower limit defined in terms of the $m_{ee}$ entry of Table \ref{tab:Constraints} with $m_3 = 0$ and $\phi_2 = \pi$.  Evaluated, this is $m_{ee}^{inverted} \approx \cos^2\theta_{13}\cos\theta_A\sqrt{\Delta m^2_A} + \mathcal{O}\left(\frac{\Delta m^2_S}{\Delta m^2_A}\right)= 0.017~{\rm eV}$.  As expected, I only find models predicting normal mass hierarchies below this bound.  Future experiments will constrain many of these symmetry structures.  Cosmological observations aimed at measuring the observable $\Sigma$ seem to have the best prospects. They have the potential to probe all of the inverted hierarchy models as well as many normal ones.  Additionally, a measurement of the neutrino hierarchy could have a great impact on flavor symmetry models.  Next generation neutrino oscillation experiments are expected
to provide non-trivial information regarding the mass spectra.
The majority are based on
neutrino/anti-neutrino asymmetries via Earth matter
effects \cite{NuRev:TheoryWhitePaper,NuRev:AndreTASI,NuRev:AndreSoWhat,MyOscHier,MyNonOscHier,AndreWinterSmallTheta13}, but depend strongly on large $\theta_{R}$ values.  The small $\theta_R$ scenario is explored in \cite{AndreWinterSmallTheta13,MyNonOscHier,MyOscHier} considering oscillation and non-oscillation searches.

It is important that these results and predictions are interpreted correctly. The nature of the fitting procedure naturally selects the lowest $\epsilon$ value allowed by the data in order to obtain the smallest possible $m_{ee}$.  With this in mind, at least one predicted parameter should sit at the edge of its allowed region for each model.  This does not mean that all of the explored symmetry structures are on the verge of exclusion.  I point out that these are only predictions for the matrix that yield the smallest $m_{ee}$ value, \emph{and} which is also consistent with data.  If future experimental constraints tighten, one must simply redo the fit with the new data.  Generally, larger symmetry breaking would be needed which would push $\epsilon$, and by extention $m_{ee}$, higher.  In terms of the histograms of Figure \ref{fig:Total4Hist}, the net result would be a general movement of $m_{ee}$ upward.  The lowest values of $m_{ee}$ are typically suppressed by the highest powers of $\epsilon$ and will be affected the most in this transformation.  It would lead to a narrower distribution.  In this process, some models may be excluded, but it is not guaranteed.

\begin{figure}
\includegraphics[scale=0.665]{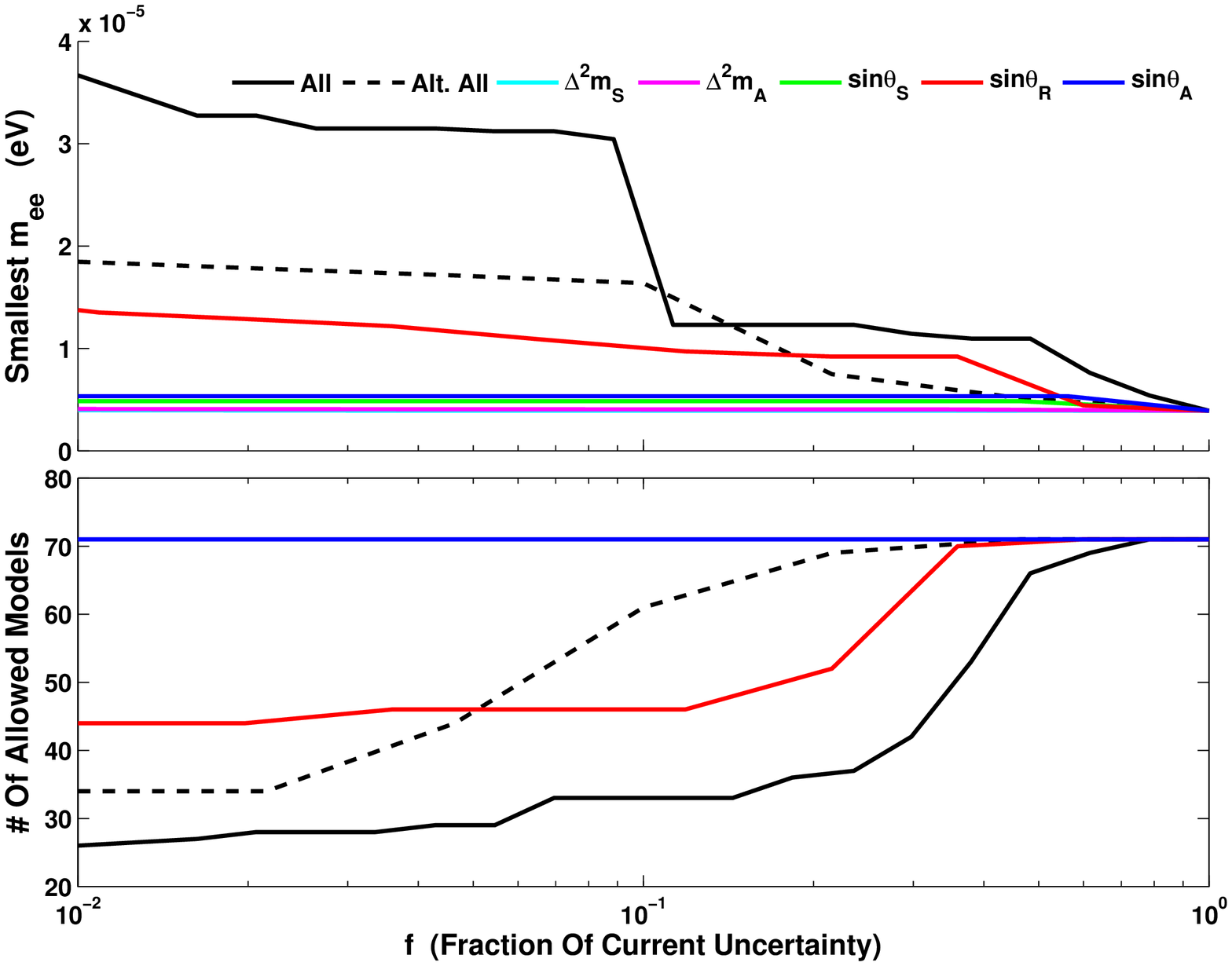}
\caption{Plot of minimization results as a function of the fractional neutrino oscillation parameter uncertainty within the $c=0.5$ case.  The upper panel shows the variation in the smallest extracted $m_{ee}$ values while the lower panel depicts the variation in the number of allowed models.  The black curves let all parameter uncertainties vary about the current best fit points (solid) and alternative central point (dashed).  This point is defined by $\sin\theta_R=0.145$ and $\sin\theta_A=0.661$ with all other parameters held at their best fit values.  This choice is consistent with data at the $68\%$ level.  For the solid colored curves, only one uncertainty is allowed to vary at a time about the best fit point with the others held at their current values.}
\label{fig:ParUncMee}
\end{figure}

In Figure \ref{fig:ParUncMee}, I show the movement of the numerical results as a function of parameter uncertainty, taken as a fraction of the current $1\sigma$ deviation, as given in Table \ref{tab:Constraints}.  Here, $f$ is defined by
\begin{equation}
 \sigma_{f}=f\sigma_{current}.
\end{equation}
The reduced $\sigma_f$ is used as input in the minimization of Eq.~(\ref{eq:Chi2}) and simulates future improvements in neutrino parameter measurements.  This is shown in a $log_{10}$ scale down to $f = 0.01$.  The black curves are obtained by varying the uncertainty on all of the oscillation parameters.  The solid and dashed lines use central values equal to the best fit parameters of Table \ref{tab:Constraints} and equal to an alternative $68\%$ allowed point, respectively.  The latter was chosen as a logical possibility with large deviations from maximal atmospheric mixing ($\sin\theta_A = 0.661$) and vanishing $\theta_R$ ($\sin\theta_R=0.145$).  All other parameters were held at the current best fit.  The solid colored curves were obtained by varying the parameter uncertainties individually, assuming the current best fit central values.  The upper and lower panels show the variation of the smallest derived $m_{ee}$ values and the total number of allowed models respectively with parameter uncertainty.  The general trend is as expected.  As the uncertainty is decreased, $m_{ee}$ is pushed larger while the number of allowed models decrease.  In terms of individual parameter curves, $\sin\theta_R$, and to a much lesser extent $\sin\theta_A$, show the largest variations.  All others induce very little deviation and almost sit directly on top of each other in the lower panel.  Thus, it is clear that improved measurements of $\theta_R$ are essential to bounding $m_{ee}$ within this framework.

The slope changing curve features of Figure \ref{fig:ParUncMee} are not numerical artifacts.  In the upper panel, they correspond to changes in the charge structure of the minimal model, each of which has its own $m_{ee}$ vs. $f$ slope.  Taking the solid black total variation curve as an example, I find that the six largest slope changes all correspond to optimal charge assignment changes.  The exact models involved here are not very enlightening.  As $f$ decreases, the optimal model class jumps from $C4$ to $C2$ to $C3$.  If the parameter best fit points remain at their current level, the bound on the minimum $m_{ee}$ value will triple, provided a two order of magnitude improvement in oscillation measurements.  Even these enhanced $m_{ee}$ values are beyond future experimental prospect, so the more significant result is related to the number of allowed models.  In this scenario, the number of allowed broken $U(1)_f$ models is reduced by roughly a factor of three to $27$.  Such a small number of possibilities would constrain neutrino mass ultraviolet completions and help guide model builders in further constructions.  Of course, this all depends on the exact neutrino parameter values. For example, the variations are not so drastic under the alternative best fit point.  This is so because the majority of the smallest $m_{ee}$ structures predict large $\theta_R$ and deviations from maximal $\theta_A$ due to symmetry breaking effects.  This point may be taken to an extreme by postulating a best fit point directly at the predictions of the minimal model; in which case, the lowest $m_{ee}$ will not vary with $f$, although the available model space might.  It is not clear if the opposite extreme exists.  Namely, is there a best fit choice that can push $m_{ee}$ into the next generation reach or that will narrow down the model space to a single, or even zero, symmetry structure?  Based on current results, the former case seems doubtful, but the latter model reduction remains a serious possibility.  This speculation may be verified by a scan over possible best fit parameter points, but such an analysis is not warranted without experimental direction in terms of improved $m_{ee}$ bounds and parameter measurements.

\section{Concluding Remarks\label{sec:Concluding-Remarks}}

While exact symmetries generating $m_{ee}=0$ are excluded by current data, slightly broken symmetry structures are still allowed that result in $m_{ee}$ well beyond the reach of future $\beta\beta0\nu$ experiments.  I systematically study these broken Abelian symmetries via a spurion analysis to obtain the smallest allowed $m_{ee}$ values.  I find that, allowing reasonable $\mathcal{O}(1)$ coupling constants, many models are excluded by data and that the smallest $m_{ee}$ is constrained to be larger than about $3.9\times 10^{-6}~{\rm eV}$ at $99\%$ confidence.  The structures yielding the smallest allowed values imply general predictions for neutrino experiments, including large deviations from vanishing $\theta_R$ and maximal $\theta_A$, in addition to the normal mass spectra.  Improvements in future data could increase these limits and perhaps, depending on the central parameter values, single out a handful of allowed models that may be explored in more depth.  Qualitatively, the main result is that there is currently a small but non-zero lower bound to $\Gamma_{\beta\beta0\nu}$ that may be improved by future precision measurements.

At face value, these results are only valid under specific circumstances.  Namely, I explore models containing three light Majorana neutrinos subject to a broken Abelian flavor symmetry with a non-trivial $\nu_e$ transformation in the absence of fine tuning.  These happen to predict small $m_{ee}$ bounds.  How far can this be pushed?  Abelian symmetries are better suited to this task than non-Abelian ones, but one may still consider discrete flavor groups.  See for example \cite{DiscreteFlavorSymmetry} and references therein.  The fact that all broken Abelian flavor models predicting $m_{ee}=0$ are excluded helps motivate that no exact Abelian discrete symmetry (a discrete subgroup of $U(1)_f$) can be allowed.  This says nothing of non-Abelian discrete groups, which is therefore a limitation of the analysis.  Next, relaxing the naturalness requirement, allowing the $a_{\alpha\beta}$ parameters to take on any value, will push down the allowed $m_{ee}$ bounds, but will not admit solutions with vanishing $\Gamma_{\beta\beta0\nu}$.  The introduction of new degrees of freedom is more complicated, but potentially testable by other means depending on the masses and couplings involved.  New physics above the $\sim 5~{\rm TeV}$ scale will decouple from the system and can be ignored for these purposes \cite{MyLNV}.  In the intermediate range between $1~{\rm GeV}$ and $5~{\rm TeV}$, heavy particle mediation of $\beta\beta0\nu$ via contact effective operators can become important and influence the relationship between $m_{ee}$ and the effective $m_{ee}^{eff}$ measured in $\beta\beta0\nu$.  These quantities should converge for small $m_{ee}$, as motivated in Section \ref{sec:introduction}, but the detailed rate depends on the new physics model.  Here, I only consider the mass matrix element, which will deviate from $m_{ee}^{eff}$ at sufficiently large $m_{ee}$.  Hence, in the presence of TeV scale new physics, the large $m_{ee}$ models of Figure \ref{fig:Total4Hist} may not be mapped onto $\Gamma_{\beta\beta0\nu}$ in the standard way.  Less caution is needed with the smallest $m_{ee}$ value models, which comprise the main object of this study.  Moving down in mass scale below approximately $100~{\rm MeV}$, the characteristic $\beta\beta0\nu$ momentum transfer, new physics can mediate the decay and interfere with the light neutrinos to suppress $\Gamma_{\beta\beta0\nu}$, regardless of the derived $m_{ee}$ value \cite{Wolfenstein:CpPropertiesMajoranaNuBB0n,MyLowESeeSaw,Andre:SeeSawEnergyScaleLSNDAnomaly,Boris:CptCpCPhasesEffectsMajoranaProcesses}.  This was discussed in \cite{MyLowESeeSaw} in terms of the eV scale type-I Seesaw mechanism, where the effect is particularly clear and can yield vanishing $\beta\beta0\nu$ rates.  In this case, for $n$ singlet neutrinos, the extended $3+n \times 3+n$ mass matrix is relevant to the $\beta\beta0\nu$ system and the upper $3\times3$ diagonal block, which includes $m_{ee}$, is zero by construction.  Thus, the relation between the light neutrino $m^{\rm light}_{ee}$ defined in the last entry of Table \ref{tab:Constraints} and $\Gamma_{\beta\beta0\nu}$ is completely spoiled.  However, one could hope to observe these light degrees of freedom in sterile neutrino oscillation searches or astrophysical phenomenon.  The last consideration is the presence of exact $m_{ee}=0$ at some high scale where neutrino masses are generated.  In the absence of a flavor symmetry to protect it, a non-zero $m_{ee}$ will be generated via two loop renormalization group effects, as motivated in \cite{SmallestNuMass}.  Details of this generation mechanism depend on new, intermediately scaled physics and has yet to be calculated for the $e-e$ element of the mass matrix.  Therefore, the qualitative existence of a lower $\Gamma_{\beta\beta0\nu}$ bound seems robust under the introduction of arbitrary new physics at scales greater than $100~{\rm MeV}$.  This offers a much needed handle, or at least a conceptual proof of principle, on means of selecting the Dirac neutrino nature.

This analysis is best suited to describe the theoretical possibilities of a scenario in the distant future where null $\beta\beta0\nu$ results push $m_{ee}$ bounds below the anarchy limit at $5\times 10^{-3}~{\rm eV}$.  Above this value, small $\Gamma_{\beta\beta0\nu}$ could be the result of mass matrix statistical fluctuations.  Thus, subject to the above testable qualifications, it would be safe to assume that one of the flavor symmetries explored here is at work to suppress $m_{ee}$ or that neutrinos are Dirac particles.  Detailed speculations on such broad experimental improvements are beyond the scope of this work, but it is reasonable that these strong $\beta\beta0\nu$ bounds would be accompanied by similar enhancements in other neutrino related parameters.  In such a world, this analysis would guide model builders toward the complete and correct neutrino mass model.  Currently, $m_{ee}$ bounds are over two orders of magnitude away from this situation.  Still, until a positive $\beta\beta0\nu$ signal is detected, this scenario remains a logical possibility that should be explored to properly understand the options open to nature.  In the meantime, experimental bounds may be used to constrain the flavor symmetry model space.

\begin{acknowledgments}
Special thanks to Andre de Gouvea and Alex Friedland for useful discussions on this topic and comments on the original manuscript. This work was performed under the auspices of the National Nuclear Security Administration of the U.S. Department of Energy at Los Alamos National Laboratory under Contract No. DE-AC52-06NA25396. This work was also funded in part by the US Department of Energy Contract No. DE-FG02-91ER40684.
\end{acknowledgments}

\bibliographystyle{apsrev}
\bibliography{/home/james/Work/MyReferences}

\end{document}